\documentclass{article}
\usepackage{subfloat}
\usepackage{lscape,epsfig}
\usepackage{graphicx}
\usepackage{amssymb}
\usepackage{amsmath}
\usepackage{natbib}
\textheight=22cm \DeclareSymbolFont{ppa}{OT1}{ppl}{m}{it}
\DeclareMathSymbol{\vv}{\mathalpha}{ppa}{'166}

minus 2.3mu \thickmuskip = 2.6mu plus 2mu minus 2.6mu

\let\svthefootnote\thefootnote

\begin{document}

\newcommand{\dd}{\,{\rm d}}
\newcommand{\ie}{{\it i.e.},\,}
\newcommand{\etal}{{\it $et$ $al$.\ }}
\newcommand{\eg}{{\it e.g.},\,}
\newcommand{\cf}{{\it cf.\ }}
\newcommand{\vs}{{\it vs.\ }}
\newcommand{\zdot}{\makebox[0pt][l]{.}}
\newcommand{\up}[1]{\ifmmode^{\rm #1}\else$^{\rm #1}$\fi}
\newcommand{\dn}[1]{\ifmmode_{\rm #1}\else$_{\rm #1}$\fi}
\newcommand{\upd}{\up{d}}
\newcommand{\uph}{\up{h}}
\newcommand{\upm}{\up{m}}
\newcommand{\ups}{\up{s}}
\newcommand{\arcd}{\ifmmode^{\circ}\else$^{\circ}$\fi}
\newcommand{\arcm}{\ifmmode{'}\else$'$\fi}
\newcommand{\arcs}{\ifmmode{''}\else$''$\fi}
\newcommand{\MS}{{\rm M}\ifmmode_{\odot}\else$_{\odot}$\fi}
\newcommand{\RS}{{\rm R}\ifmmode_{\odot}\else$_{\odot}$\fi}
\newcommand{\LS}{{\rm L}\ifmmode_{\odot}\else$_{\odot}$\fi}

\newcommand{\Abstract}[2]{{\footnotesize\begin{center}ABSTRACT\end{center}
\vspace{1mm}\par#1\par \noindent {~}{\it #2}}}

\newcommand{\TabCap}[2]{\begin{center}\parbox[t]{#1}{\begin{center}
  \small {\spaceskip 2pt plus 1pt minus 1pt T a b l e}
  \refstepcounter{table}\thetable \\[2mm]
  \footnotesize #2 \end{center}}\end{center}}

\newcommand{\TableSep}[2]{\begin{table}[p]\vspace{#1}
\TabCap{#2}\end{table}}

\newcommand{\FigCap}[1]{\footnotesize\par\noindent Fig.\  %
  \refstepcounter{figure}\thefigure. #1\par}

\newcommand{\TableFont}{\footnotesize}
\newcommand{\TableFontIt}{\ttit}
\newcommand{\SetTableFont}[1]{\renewcommand{\TableFont}{#1}}
\newcommand{\MakeTable}[4]{\begin{table}[htb]\TabCap{#2}{#3}
  \begin{center} \TableFont \begin{tabular}{#1} #4
  \end{tabular}\end{center}\end{table}}

\newcommand{\MakeTableSep}[4]{\begin{table}[p]\TabCap{#2}{#3}
  \begin{center} \TableFont \begin{tabular}{#1} #4
  \end{tabular}\end{center}\end{table}}

\newenvironment{references}%
{ \footnotesize \frenchspacing
\renewcommand{\thesection}{}
\renewcommand{\in}{{\rm in }}
\renewcommand{\AA}{Astron.\ Astrophys.}
\newcommand{\AAS}{Astron.~Astrophys.~Suppl.~Ser.}
\newcommand{\ApJ}{Astrophys.\ J.}
\newcommand{\ApJS}{Astrophys.\ J.~Suppl.~Ser.}
\newcommand{\ApJL}{Astrophys.\ J.~Letters}
\newcommand{\AJ}{Astron.\ J.}
\newcommand{\IBVS}{IBVS}
\newcommand{\PASP}{P.A.S.P.}
\newcommand{\Acta}{Acta Astron.}
\newcommand{\MNRAS}{MNRAS}
\renewcommand{\and}{{\rm and }}
\section{{\rm REFERENCES}}
\sloppy \hyphenpenalty10000
\begin{list}{}{\leftmargin1cm\listparindent-1cm
\itemindent\listparindent\parsep0pt\itemsep0pt}}%
{\end{list}\vspace{2mm}}

\def\TYLDA{~}
\newlength{\DW}
\settowidth{\DW}{0}
\newcommand{\dw}{\hspace{\DW}}

\newcommand{\refitem}[5]{\item[]{#1} #2%
\def\REFARG{#3}\ifx\REFARG\TYLDA\else, {\it#3}\fi
\def\REFARG{#4}\ifx\REFARG\TYLDA\else, {\bf#4}\fi
\def\REFARG{#5}\ifx\REFARG\TYLDA\else, {#5}\fi.}

\newcommand{\Section}[1]{\section{#1}}
\newcommand{\Subsection}[1]{\subsection{#1}}
\newcommand{\Acknow}[1]{\par\vspace{5mm}{\bf Acknowledgements.} #1}
\pagestyle{myheadings}

\newfont{\bb}{ptmbi8t at 12pt}
\newcommand{\xrule}{\rule{0pt}{2.5ex}}
\newcommand{\xxrule}{\rule[-1.8ex]{0pt}{4.5ex}}

\begin{center}
{\Large\bf
 The Clusters AgeS Experiment (CASE).  \\
 Variable stars in the field of  
 the globular cluster M12}{\LARGE$^\ast$}
 \vskip1cm
  {\large
      ~~J.~~K~a~l~u~z~n~y$^1$\dag,
      ~~I.~B.~~T~h~o~m~p~s~o~n$^2$,
      ~~W.~~Narloch$^1$
      ~~W.~~P~y~c~h$^1$
      ~~and~~M.~~R~o~z~y~c~z~k~a$^1$
   }
  \vskip3mm
{ $^1$Nicolaus Copernicus Astronomical Center, ul. Bartycka 18, 00-716 Warsaw, Poland\\
     e-mail: (jka, wnarloch, pych, mnr)@camk.edu.pl\\
  $^2$The Observatories of the Carnegie Institution for Science, 813 Santa Barbara
      Street, Pasadena, CA 91101, USA\\
     e-mail: ian@obs.carnegiescience.edu}
\end{center}

\vspace*{7pt}
\Abstract 
{The field of the globular cluster M12 (NGC 6218) was monitored between 
1995 and 2009 in a search for variable stars. $BV$ light curves were obtained 
for 36 periodic or likely periodic variables. Thirty-four of these are new detections. 
Among the latter we identified 20 proper-motion members of the cluster: six 
detached or semi-detached eclipsing binaries, five contact binaries, five 
SX~Phe pulsators, and three yellow stragglers. Two 
of the eclipsing binaries are located in the turnoff region, one on the lower 
main sequence and the remaining three among the blue stragglers. Two contact systems 
are blue stragglers, and the remaining three reside in the turnoff region.  
In the blue straggler region a total of 103 objects were found, of which 42 are 
proper motion members of M12, and another four are field stars. Forty-five of the remaining 
objects are located within two core radii from the center of the cluster, and as 
such they are likely genuine blue stragglers. We also report the discoveries of a
radial color gradient of M12, and the shortest period among contact systems 
in globular clusters in general.
}
{globular clusters: individual (M12) -- stars: variables -- 
stars: SX Phe -- blue stragglers -- binaries: eclipsing
}

\let\thefootnote\relax\footnotetext{\dag Deceased}
\let\thefootnote\relax\footnotetext
{$^{\mathrm{\ast}}$Based on data obtained with du Pont and Swope telescopes at Las
 Campanas Observatory.}
\let\thefootnote\svthefootnote

\Section{Introduction} 
\label{sec:intro}
M12 is a nearby ($(m-M)_{V}=14.01$ mag) globular cluster located at 
a high galactic latitude ($b=26.3$~deg) in a field of low reddening 
with $E(B-V)=0.19$ mag. According to Harris 1996 (2010 edition), its 
core radius, half-light radius, [Fe/H] index and radial velocity are equal 
to 0$'$.79, 1$'$.77, -1.37 and -41.4$\pm$0.2 km/s, respectively. 
Low reddening and relatively low concentration make it
an attractive target for detailed studies with ground-based telescopes.
The photometric survey presented here is a part of the CASE project (Kaluzny 
et al. 2005) conducted with telescopes of the Las Campanas Observatory, Chile.

Early pre-CCD searches for variable stars in the field of M12 were summarized 
by Clement et al. (2001). These studies resulted in the detection of just one variable 
- a bright, long-period pulsator of W Vir type, which is overexposed in our 
frames. Based on CCD data, von Braun et al. (2002) 
found another two variables, both of W UMa type (we recovered these, and 
we retain their original names V1 and V2). To the best of our 
knowledge, no additional discoveries have been reported. 

In this contribution we present results of a long-term photometric survey
conducted between 1995 and 2009.
Section 2 contains a report on the observations and explains the
methods used to calibrate the photometry. The detected variables
are presented and discussed in Section~3. The paper is summarized 
in Section 4.
\section{Observations}
\label{sec:obs}
Our paper is based on two sets of images. The first set was obtained using 
the 2.5-m du Pont telescope and the $2048\times 2048$ TEK5 CCD camera with
a field of view of 8.84 arcmin on a side at a scale of 0.259
arcsec/pixel. Observations were conducted on 23 nights from April 
26, 1999 to June 29, 2009. The same set of filters was used for all observations. 
For the analysis, we used 769 $V$-band images with seeing 
ranging from 0$''$.64 to 2$''$.07, and 216 $B$-band images with seeing 
ranging from 0$''$.71 to 2$''$.13. The median value of the seeing was 
1$''$.07 and 1$''$.14 for $V$ and $B$, respectively. Additionally, a few $U$-band 
frames were taken in April/May 2001 at an average seeing of 1$''$.25. 
The second set of images was obtained with the 1.0-m Swope telescope 
using the $2048\times 3150$ SITE3 camera.
The field of view was $14.8\times 22.8$ arcmin$^2$ at a scale of 0.435
arcsec/pixel. About 50\% of the images were taken with a subraster
providing a field of view of $14.8\times 14.8$ arcmin$^2$.
Observations were conducted on 57 nights from April 15, 1999 to June
08, 2008. Again, the same filters were used for all observations.
For the analysis, we used 978 $V$-band images with seeing ranging from 0$''$.95 
to 2$''$.31 and 168 $B$-band images with seeing ranging from 1$''$.03 to 2$''$.15. 
The median value of the seeing was 1$''$.37 and 1$''$.45 for $V$ 
and $B$, respectively. 

\begin{figure}
   \centerline{\includegraphics[width=0.95\textwidth,
               bb = 9 11 719 399, clip]{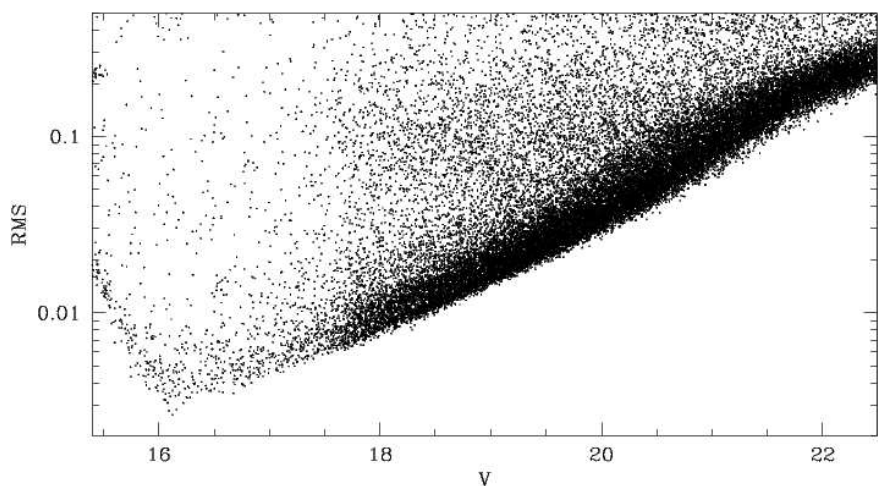}}
   \caption{ Standard deviation vs. average $V$ magnitude for
    light curves of stars from the M12 field. Light curves
    are based on images from the du Pont telescope.
    \label{fig:rms}}
\end{figure}

The photometry was performed using an image subtraction technique.
The du Pont data were reduced with a modified version of the ISIS V2.1 package 
(Alard  2000), whereas for the frames obtained with the Swope telescope the DIAPL 
package\footnote{Available from http://users.camk.edu.pl/pych/DIAPL/index.html} 
was used. For each set and each filter, a reference image was constructed
by combining several high quality frames. Daophot, Allstar and Daogrow codes 
(Stetson 1987, 1990) were used to extract the profile photometry, and to
derive aperture corrections for the reference images.
Additionally, profile photometry was extracted for individual images from
the du Pont telescope. This allowed us to obtain useful measurements for
stars which were overexposed on the reference images. Also,
profile photometry enabled an unambiguous identification of variable
stars in crowded fields, which is sometimes problematic when image 
subtraction  only is used. We attempted to resolve numerous blends in 
the Swope data for the central part of M12 by using star positions from du Pont 
photometry. In most cases this approach proved to be successful.
The accuracy of the du~Pont photometry is illustrated in Fig. \ref{fig:rms},
in which the standard deviation of the photometric measurements is plotted 
as a function of the average magnitude in~$V$. 
\subsection{Calibration}

\begin{figure}
   \centerline{\includegraphics[width=0.95\textwidth,
               bb = 58 314 562 687, clip]{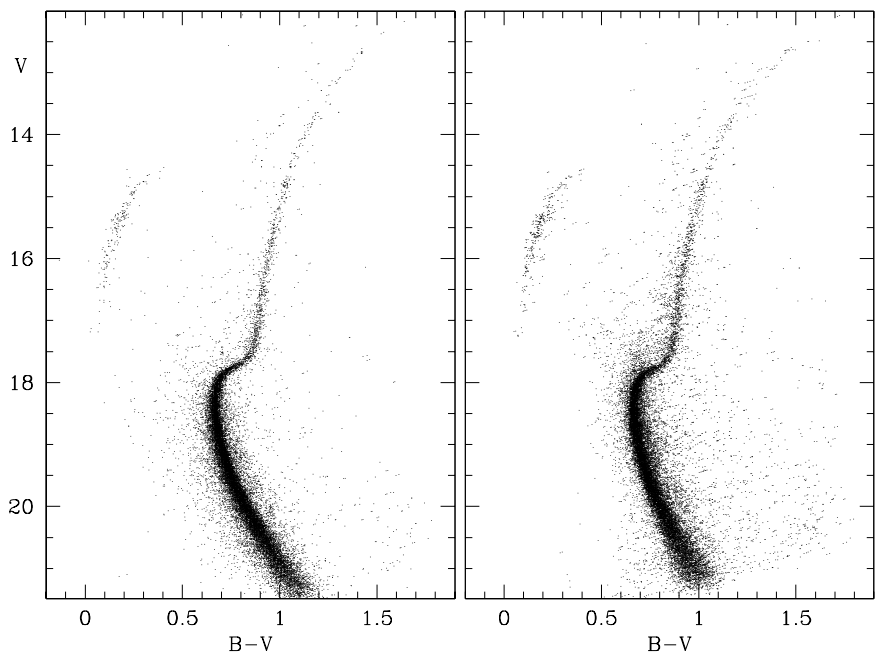}}
   \caption{CMDs of M12 based on the data from the du Pont telescope 
    (left) and the Swope telescope (right).
    \label{fig:cmds}}
\end{figure}

The photometry collected with the du Pont telescope was transformed to the
standard $UBV$ system based on observations of stars from
Landolt fields (Landolt 1992). On the night of May 29, 2001 we
observed 35 stars from five such fields.
These data were used to find the coefficients of linear transformation 
from the instrumental system to the standard one. Residual differences 
between the standard and recovered magnitudes and colors amounted to
0.008, 0.007 and 0.009 mag for $V$, $B$ and $B-V$, respectively. 
The residuals did not show any systematic dependence on the color index.
Transformations for the photometry obtained with the Swope telescope were
based on the calibrated data from the du Pont telescope.
Linear transformations proved to be entirely adequate.
Fig.~\ref{fig:cmds}, based on reference images, shows the color-magnitude 
diagram (CMD) of the observed fields, illustrating the range of stellar 
population examined for variability (stars with formal error in 
$V$ larger than 0.05 mag and formal error in $B-V$ larger than 0.1 mag 
are not shown). The contamination of the cluster by field 
interlopers is rather low, but clearly increases with the larger 
field of view of the Swope data. Full sets of photometry can be downloaded from the CASE 
archive.\footnote{http://case.camk.edu.pl/} 
\subsection{Search for variables}
The search for variable stars was conducted using the AOV and AOVTRANS
algorithms implemented in the TATRY code (Schwarzenberg-Czerny
1996, Schwar\-zenberg-Czerny \& Beaulieu 2006). We examined the
du Pont light curves of 27586 stars with $V<21.5$ mag and the Swope 
light curves of 23265 stars with $V<20$ mag. The limits of detectable 
variability depended on the accuracy of photometric measurements, which
for the du Pont data decreased from 3~mmag at $V = 16$~mag to 30~mmag at $V=20$~mag 
and 100~mmag at $V=21.5$~mag (Fig.~\ref{fig:rms}). For the Swope data 
the accuracy decreased 
from 5~mmag at $V=15$~mag to 50~mmag at $V=19$~mag and 100~mmag at 
$V=20$ mag.

\section{The variables}

\begin{figure}
   \centerline{\includegraphics[width=0.95\textwidth,
               bb = 101 192 509 600, clip]{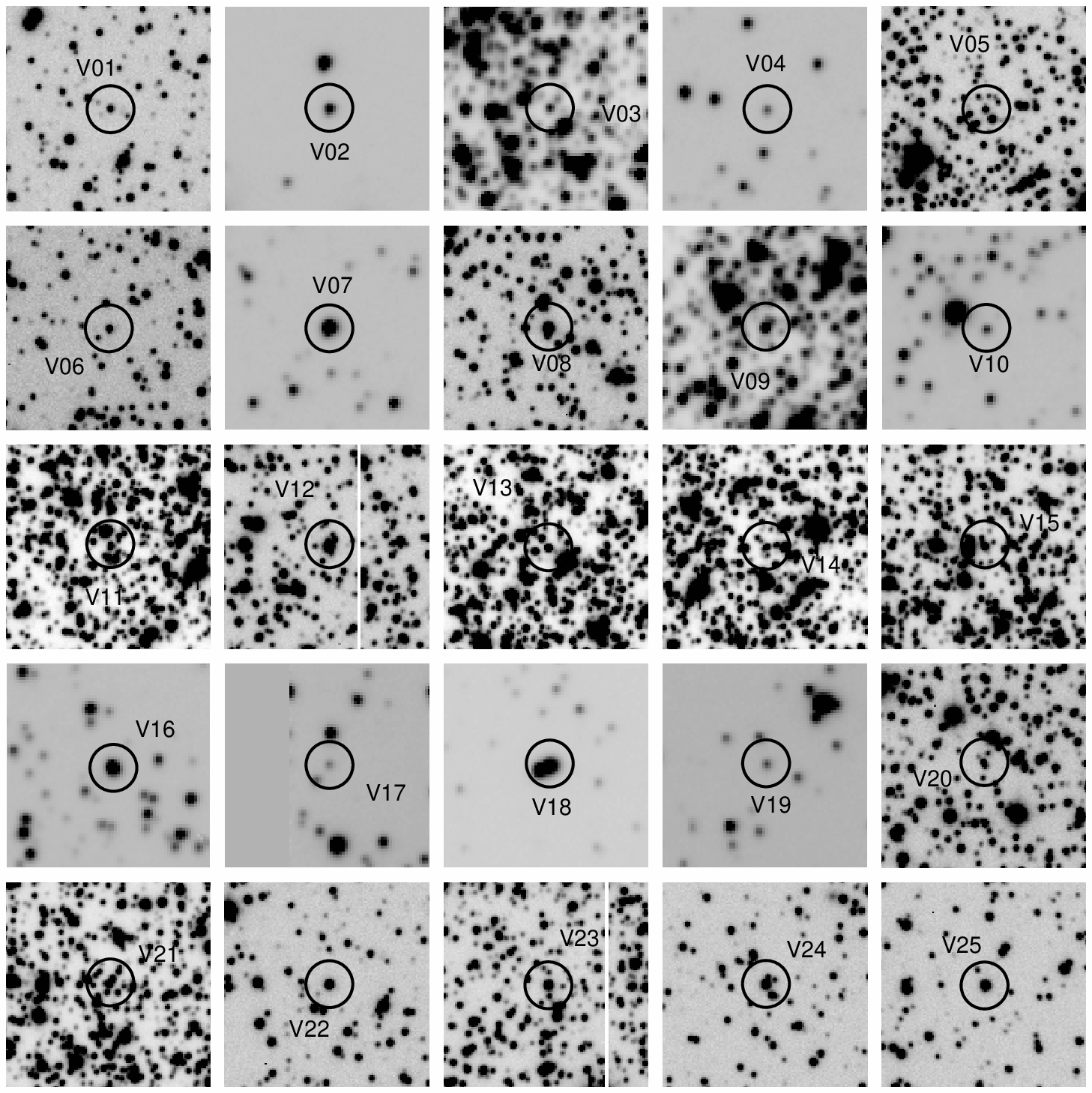}}
   \vspace{1.5 mm}
   \centerline{\includegraphics[width=0.95\textwidth,
               bb = 101 358 509 600, clip]{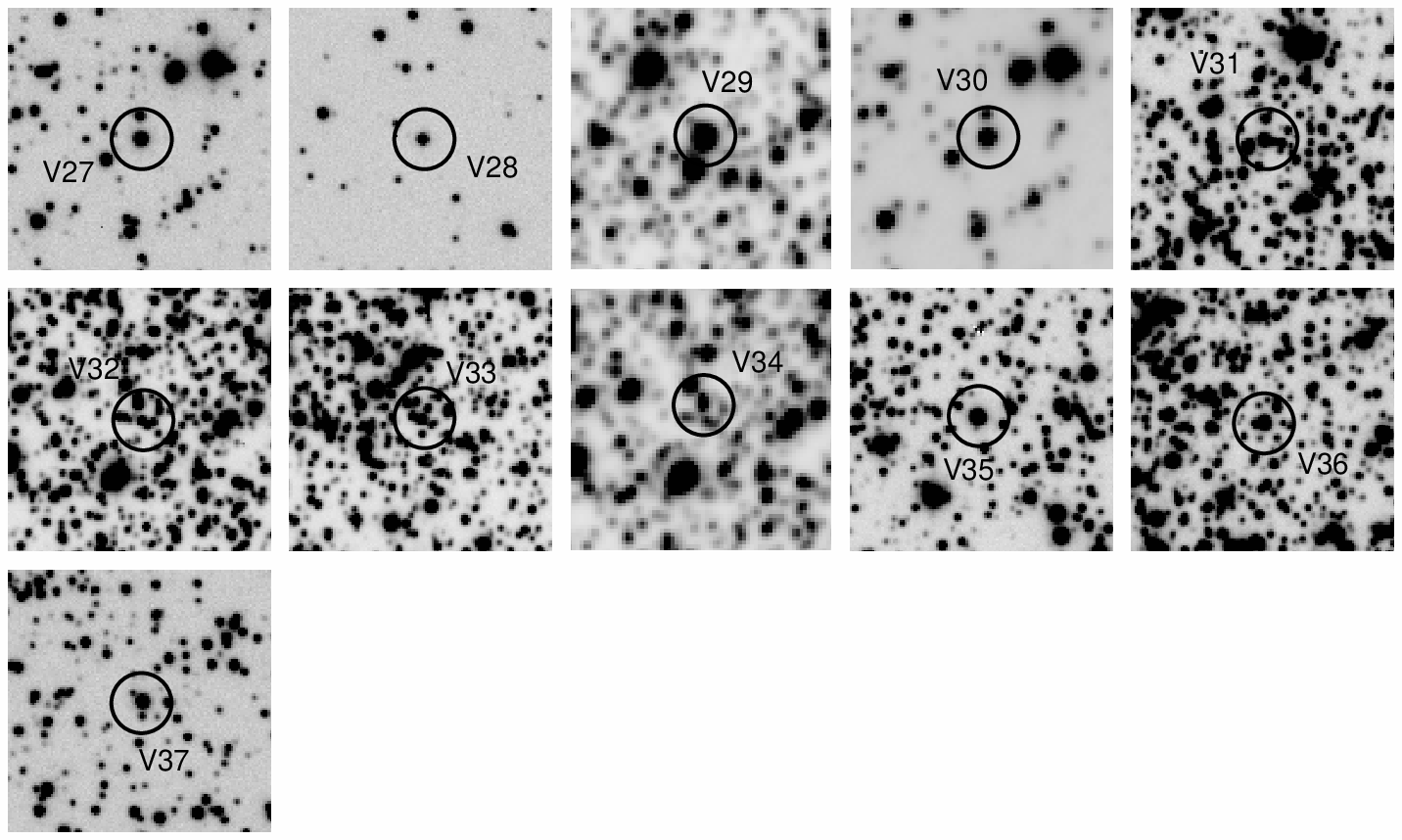}}
\caption{Finding charts for the variables.
    Each chart is 30 arcsec on a side; north is up and east to the left.
    \label{fig:maps}}
\end{figure}

We detected 33 certain and three suspect variables, of which 20 have photometry
from both telescopes. Only two of these had been known previously, the two W UMa
systems discovered by von Braun et al. (2002). Fig.~\ref{fig:maps} presents finding charts 
for all 36 variables. Basic properties of these variables are listed in Table 1. 
The equatorial coordinates in Table 1 conform to the UCAC4 system (Zacharias et al. 2013),
and are accurate to about 0$''$.2. We checked that none of our variables coincides with 
any X-ray source from the list of Lu et al. (2009) or with the UV source discovered by Schiavon
et al. (2012). 

The $V$-magnitudes listed in Table 1 correspond to the maximum light in the 
case of eclipsing binaries, while for the remaining variables average magnitudes
are provided. For each variable the $B-V$ color is given, followed by the amplitude 
in the $V$-band. Periods of variability were found for all stars except two eclipsing 
binaries, for which parts of single eclipses only were observed. Some light curves 
show phase shifts and/or change shape from season to season. In these cases we give periods 
obtained for the indicated season. The last column of Table 1 gives the membership status 
based on proper motions (PM) taken from Zloczewski et al. (2012) and Narloch 
et al. (in preparation). Phased light curves of the variables from Table 1 are presented 
in Figs.~\ref{fig:curves1}, \ref{fig:curves2} and \ref{fig:curves3}. 

\begin{figure}
   \centerline{\includegraphics[width=0.95\textwidth,
               bb = 43 170 562 688, clip]{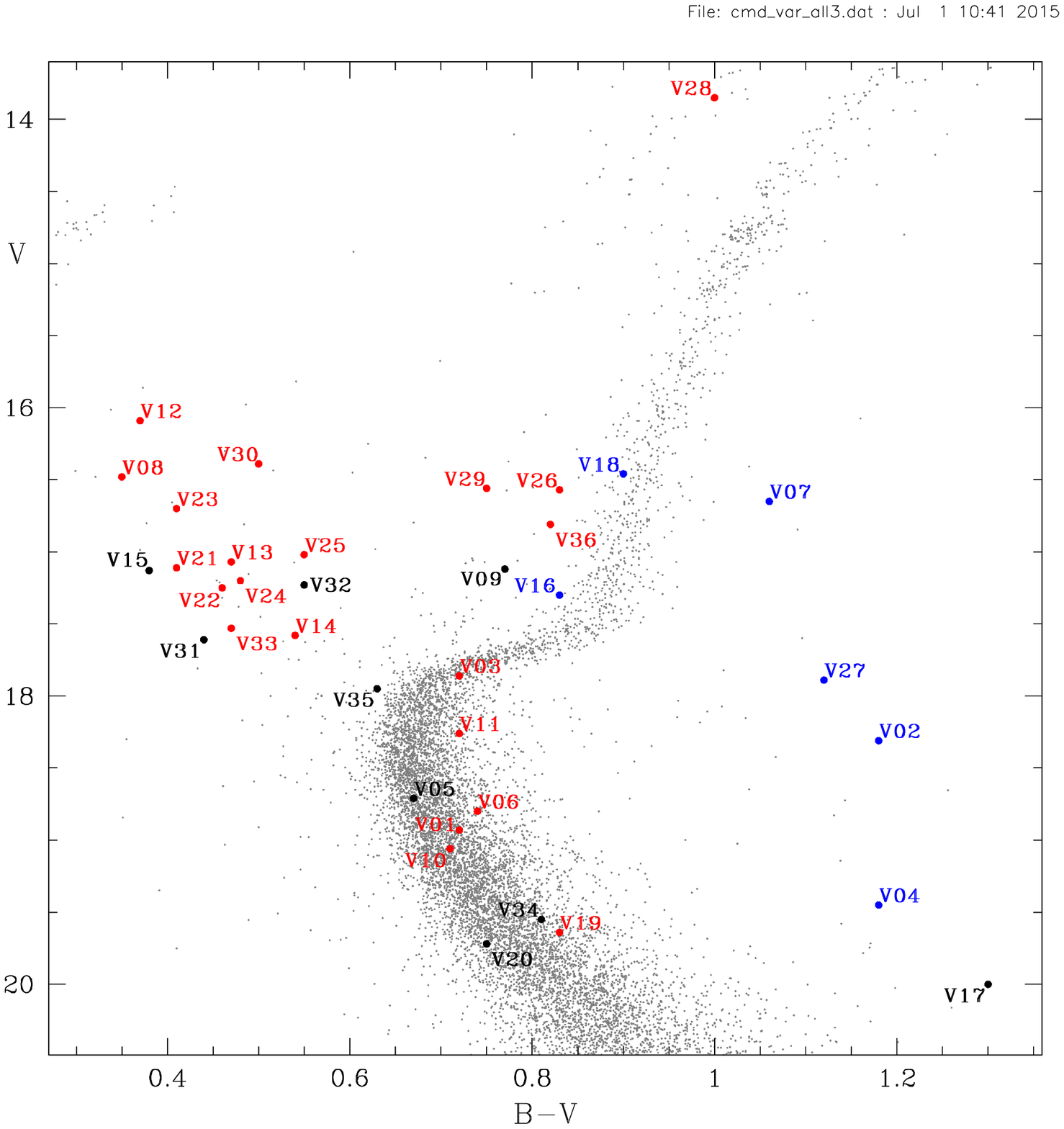}}
   \caption{Color-magnitude diagram for M12 with indicated locations of 
    the variables. Red, blue and black labels denote, respectively, members, nonmembers
    and objects for which the membership data is missing or ambiguous.
    \label{fig:cmd_var}}
\end{figure}

A CMD of the cluster with the locations of the variables is shown in Fig.~\ref{fig:cmd_var}.  
This diagram is based on the du Pont photometry, and it only includes stars classified by 
Zloczewski et al (2012) as PM members of M12. Variables that are PM members of M12 are labelled 
in red, those with PM indicating that they are field objects  in blue, and those 
for which the PM data are missing or ambiguous  in black. The exception is V11 for which 
we have no PM data. We attribute to it a legitimate cluster membership based on our as yet 
unpublished radial velocity measurements. 

\subsection{Detached eclipsing binaries}
\label{subs:EA}

We detected 11 detached eclipsing binaries, of which six are proper motion 
members of the cluster. Systems V10 and V11 are located at the turnoff region. With orbital 
periods amounting to 4.6 d and 5.2 d, respectively, they are interesting targets for a detailed 
follow up study aiming at the determination of their absolute parameters, and the age and 
distance of M12. We are presently conducting such an analysis for V11. The systemic velocity 
of this binary, equal to -44.4 km s$^{-1}$, differs by less than 10\% from the systemic velocity 
of M12 listed by Harris (1996, 2010 edition). This, 
together with its location at a distance of $15''$ (i.e. 0.32 core radii) from the center of 
the cluster leaves little doubt about its membership. It is somewhat 
surprising that V11 has a markedly eccentric orbit ($\epsilon\approx0.1$) despite a 
relatively short period. At an age above 13 Gyr (Dotter et al. 2010), it should have been 
fully circularized by tidal friction (Mazeh 2008; Mathieu et al. 2004). Since 
we found no indication for a third body in this system, we speculate that the orbit of 
V11 was distorted during the last few Gyr as a result of a close encounter.

The light curve of V10 shows two partial eclipses of similar depth ($\sim0.35$ mag 
in $V$). As a result, the photometric solution is likely to be degenerate, allowing for 
a broad range of relative radii. Such a degeneracy may be overcome by the determination 
of the light ratio from spectra, but this will be difficult given the 
faintness of the system. A much deeper eclipse ($\sim0.95$ mag in $V$) was partly observed 
in V19. As this binary is still fainter than V10, a really 
large observational effort would be required to collect enough data for an accurate analysis. 
The same argument concerns V20, for which an eclipse at least 0.6 mag deep in $V$ was partly
observed. The PM-membership is unclear for this binary, however its location on the main 
sequence of the cluster and at the edge of the cluster's core (62$''$ from the center) 
both suggest that it does  belong to M12.  

A potentially very interesting object is the eclipsing binary V16. We marked it as a nonmember 
in Fig.~\ref{fig:cmd_var}, however it almost meets the membership criterion employed by 
Zloczewski et al. (2012). It is located on the lower red giant branch of M12, 3$'$.8 (i.e. 
$\sim$2 half-light radii) away from the center of the cluster. Since we observed only three 
largely incomplete eclipses, the period of 2.5 d given in Table 1 is a tentative one only, and the 
eccentricity visible in Fig. \ref{fig:curves2} may be spurious. If this bright and well-isolated 
system turns out to be a member of M12, it will provide tight limits for age and distance 
of the cluster. V18 seems to be even more promising: based on its location on the 
red giant branch one might expect it to deliver more precise data than V16 which is by $\sim$1~mag 
fainter. Unfortunately, the proper motion of this system proves that we are dealing with a nearby 
binary which quickly ($\sim$14 mas yr$^{-1}$) moves across the M12 field. 
The PM-membership status of V17 is ambiguous, but its location far to the right of the 
unevolved main sequence of M12 indicates that it is a field object - probably a pair of 
nearby red dwarfs. 

Variables V12, V13, V14, V15 are located in the blue straggler region between the turnoff and the 
horizontal branch. The first three of these are cluster members with periods of 1.03, 0.73 and 
0.46 d, respectively, and light curves ranging from EA with proximity effect in V12 to EB-like 
in V14. Well defined moments of contact, together with  straight ingress and egress branches of 
the primary minimum, suggest that they are detached rather than semi-detached. The PM-data for V15 
are ambiguous: Zloczewski 
et al (2012) classify it as a probable member, while Narloch et al. (in preparation) as a 
nonmember. However, both its position on the CMD and its location in the field of view 
(24$''$, i.e. 0.3 core radii from the center of M12) suggest that this system also belongs 
to the cluster. An additional support for the membership of V15 comes from its nature. 
Short period (0.54 d), moderate proximity effect and narrow total eclipses with different 
depths indicate a system with low mass ratio and small secondary which is much hotter 
than the primary (thus, in the present configuration of the system the main minimum is due 
to the eclipse of the secondary). Preliminary 
modeling with the PHOEBE facility (Pr\v{s}a and Zwitter 2005) yields $q\approx0.1$, 
$T_p\approx7500$ K, $T_s\approx12000$ K, and $R_s/R_p\approx0.15$. Apparently, the secondary has 
lost most of its H-rich envelope to the primary, thus conforming to one of the scenarios of the 
origin of blue straggler stars. 

The secondary minima of V12, V13 and V14 are 
much shallower than the primary minima, likely because of low mass ratios. 
We suggest that these too have experienced significant mass-transfer episodes. A detailed study of 
the whole quartet would certainly deliver valuable information concerning the nature and evolution 
of blue straggler stars. 

\subsection {Contact binaries}
\label{subs:EW}

We identified nine variables with W UMa-type lightcurves, of which four are PM-members of 
the cluster, and another three are field interlopers. PM data for the remaining two 
systems are absent or not accurate enough to evaluate  membership probability. 
Small amplitudes of V08 and V09 suggest that we may deal with elliptical variables rather than 
genuine contact binaries. Spectroscopical data are needed to resolve this ambiguity.
Kaluzny et al. (2014) found a general paucity of contact systems on the unevolved main 
sequences of globular clusters. M12 perfectly conforms to this 
rule: PM-members V01, V03 and V06 and the suspected PM-member V05 reside in the turnoff 
region while another PM-member, V08, is a blue straggler. Thus we add to the growing
evidence that, at least in globular clusters, the principal factor enabling contact 
systems to form from close but detached binaries is nuclear evolution: a contact
configuration is achieved once the more massive component starts to expand quickly 
at the turnoff. Apparently, nuclear evolution is more important in this respect 
than the frequently invoked magnetic breaking; see e.g. Stepien \& Gazeas (2012) and 
references therein. 

The binary V09 is a yellow straggler candidate. Based on the available data, its PM-membership 
cannot be firmly established. The empirical calibration of Rucinski (2000) yields a distance 
modulus by 0.7 mag smaller than that given by Harris (1996, 2010 edition)for M12. Given the large 
spread of the calibration, this is also an ambiguous result. However, V09 resides only 42$''$ 
away (0.6 core radii) from the center of M12, and it may well be a cluster member. Since yellow 
stragglers are rather rare objects, it deserves further study aimed at clarifying its membership 
and evolutionary status. We note in passing that the periods of all abovementioned W UMa stars except 
V08 and V09 are shorter than 0.26 d, i.e. they are exceptionally short even for contact systems 
in globular clusters (Rucinski 2008). Among these, V03 with $P=0.210636$ d seems to be the new 
record holder.

\subsection{Variable stars among blue stragglers}
The CMD based on the du Pont data contains 103 candidate blue stragglers with $16.0<V<17.7$ and 
$0.23<B-V<0.70$. Of these 42 are proper motion members of M12, four are field stars, and for 57 
objects no PM data are available. Of the latter 45 are located within two core radii from the 
center of the cluster and are  likely members. In addition to the five eclipsing binaries described 
in Subsections \ref{subs:EA} and \ref{subs:EW}, nine variables 
were found among the stragglers, seven of which are confirmed and another two suspected members of 
the cluster. Five of these are SX Phe pulsators with periods ranging from 0.019 to 0.049 d. The star 
with the longest period, V25, has the largest amplitude ($\sim$0.2 mag in $V$) and a light curve 
characteristic of multimodal pulsations. Unfortunately, the photometry is not good enough to 
enable a period analysis. Stars V30 -- V33 exhibit more or less sinusoidal luminosity variations 
whose nature is difficult to establish. Doubling their periods makes their light curves resemble 
those of W~UMa variables. However in that case the light curves of V30 and V31 become 
less regular, whereas for V32 and V33 the periods themselves grow too long for contact 
binaries (Rucinski 2007). 
\subsection{Remaining objects}

V26, V29 and V36 are members of M12 located in the yellow straggler region, and as such they 
should be paid particular attention in future studies. V26 exhibits sinusoidal variations with 
$P=1.84$ d whose phase and ampliture vary from season to season. Fig. \ref{fig:curves2} shows 
data from our best season (2001), in which almost 650 frames were collected during 16 nights. 
V29 showed clear sinusoidal variation in 2001 only, with two possible periods of 0.647 and 1.847~d.
V36 is a suspected variable with a very noisy light curve and amplitude of only $\sim0.02$ mag. 
Another M12 member, V28, is located slightly below the asymptotic giant branch. Regular variations, 
perhaps the due to spots on the surface of a faint companion, were observed in 2001 only. 
The field object V27 is a flare star exhibiting sinusoidal variations with $P=11.4$ d 
whose phase amplitude and average magnitude vary from season to season. Before 2001 the star 
faded from 17.75 mag to 18.15 mag in $V$. On June 3, 2001 a complete flare was observed: the brightness 
smoothly increased by 0.2 mag in 1.2 h, and within the next 4 h it returned to the pre-flare 
level. Five days later regular variations were again dominating, with mean brightness 17.9 mag, 
amplitude $\sim0.3$ mag, and the same period as before the flare. The flare itself is 
visible in Fig. \ref{fig:curves2} as a thin spike at phase 0.4. The nature of the periodic variation
is unclear - spectroscopic observations are necessary to discriminate between rotation and orbital
motion of V27. The membership of suspected variables V34 and V35 is unclear. Both these objects seem 
to exhibit roughly sinusoidal variations with $P=0.35$ d and $P=0.96$ d, respectively.

\Section{Summary}
 \label{sec:sum}
We have conducted an extensive photometric survey of the globular cluster M12 in a search 
for variable stars. A total of 31 variables plus three suspected variables were discovered, and 
multiseasonal light curves were compiled for another two W UMa type eclipsing binaries that had been 
known before. For all variables periods accurate to 0.00001 -- 0.001 d were obtained. 
Seven eclipsing binaries and five pulsating stars (all of them PM-members of the cluster)
were found in the blue-straggler region. Four of the binaries are likely to have low 
mass ratios, most probably due to mass-transfer episodes. Their detailed analysis should
deliver very valuable information concerning nature and evolution of blue stragglers.
Yellow stragglers are represented in our sample by three confirmed and one suspected 
member of the cluster. The latter is a W UMa system and the nature of the former 
is unclear. Since yellow stragglers are even more rare and interesting than the blue
stragglers, these two systems should be paid close attention during future observations. Two 
detached eclipsing binaries belonging to the cluster were identified in the turnoff 
region, and another one at the lower main sequence. A potentially very valuable discovery 
is the detached eclipsing binary V16 residing on the lower red giant branch. If this bright 
and well isolated object turns out to be a member of M12, it will provide tight limits 
on the  age and distance of the cluster. 

\Acknow
{JK, WN, WP and MR were partly supported by the grant DEC-2012/05/B/ST9/03931
from the Polish National Science Center. 
}

\clearpage

\section*{Appendix: Color gradient}

\begin{figure}
   \centerline{\includegraphics[width=0.95\textwidth,
               bb = 43 175 562 688, clip]{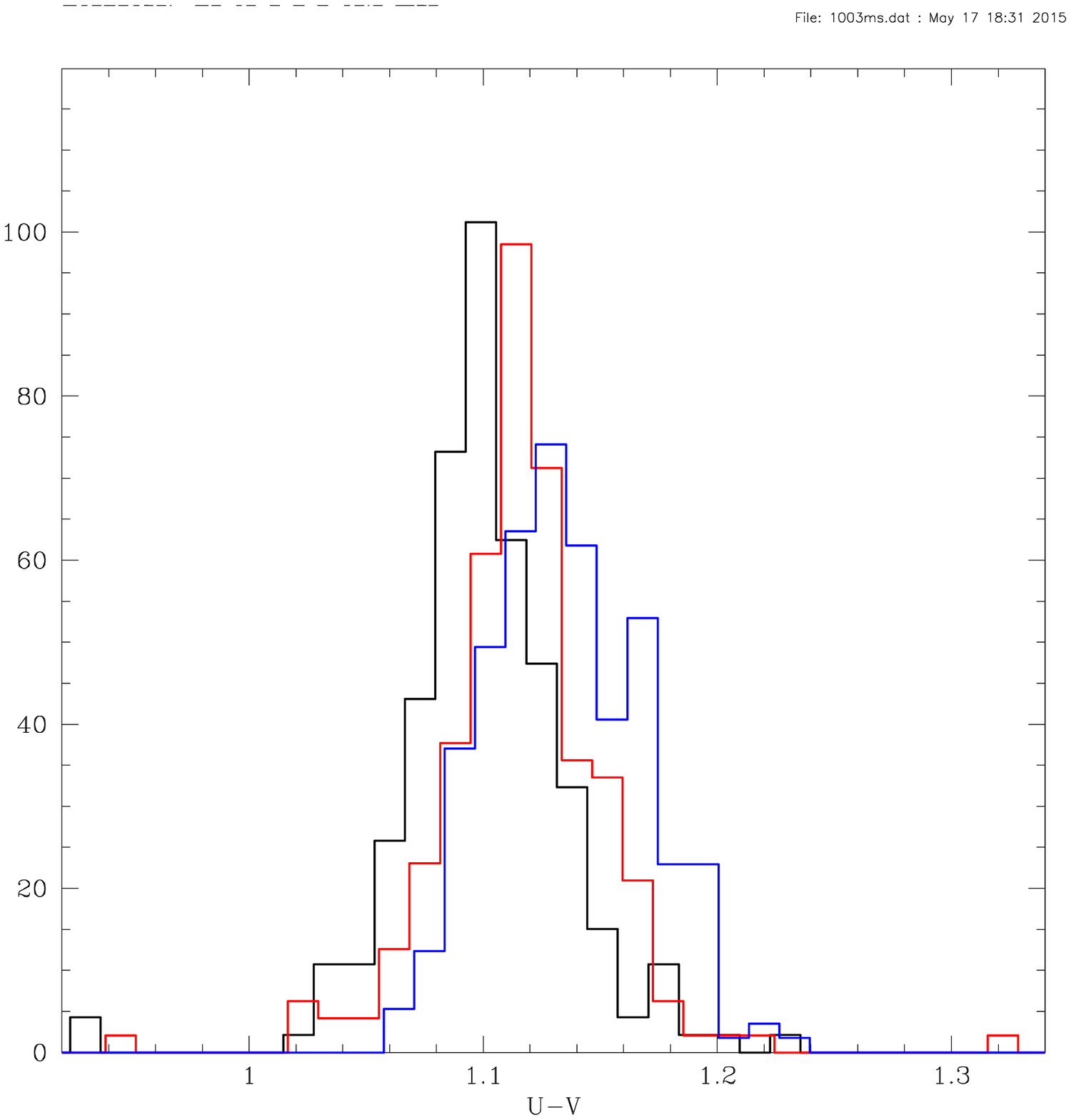}}
   \caption{Histogram of $U-V$ indices of M12 PM-members. Main sequence stars with $18<V<19.5$ 
    mag are only shown, and the normalization is arbitrary. Black: central circle $0\le r\le 2'.36$;
    red: intermediate ring $2'.36<r<3'.74$; blue: outer region $r\ge 3'.74$.
    \label{fig:umv_hist}}
\end{figure}

In some GC color gradients have been observed. Roediger et al. (2014) 
listed six such clusters. Since that sample does not include M12, we thought it worthwhile 
to search for radial variations of the colors 
in our data. The search was limited to du Pont frames, a few of which were taken in $U$ band. 
Only PM-members of M12 with $V<19.5$ mag and color errors smaller than 0.015 mag were included. 
The field of view was divided into three concentric subfields centered on the center of the cluster,
with $0\le r\le 2'.36$, $2'.36<r<3'.74$ and $r\ge 3'.74$, respectively. Each subfield contained almost 500 
stars. Weighted means of $\overline{B-V}$ and $\overline{U-V}$ colors were calculated separately 
for giants and HB stars $(V<17.5$ mag), subgiants  $(17.5\le V <18.0$ mag), and dwarfs 
$(V\ge18$ mag). No clear dependence of $\overline{B-V}$ of $r$ was found in any of the luminosity classes. 
The same negative result was obtained for $\overline{U-V}$ for  giants or subgiants. A clear, albeit 
weak, dependence emerged only in the case of the $U-V$ color for main sequence stars, with $\overline{U-V}$
increasing from $1.1002\pm0.0009$ mag in the central circle through $1.1142\pm 0.0009$ mag in the intermediate
ring to $1.1331\pm 0.0008$ mag in the outer region. This radial trend is convincingly illustrated by the 
histograms shown in Fig. \ref{fig:umv_hist}. We attribute this result at least partly to the relative underabundance 
of faint red main-sequence stars discovered by De Marchi et al. (2007) near the cluster center. 

\begin{table}
\footnotesize
 \begin{center}
 \caption{Basic data of M12 variables identified within the present survey$^*$
          \label{tab:vardata}}
 \begin{tabular}{|c|c|c|c|c|c|c|l|c|}
  \hline
 Id & RA & DEC & $V$ & $B-V$ & $\Delta V$& Period  & Remarks$^a$& Mem$^b$ \\
    & [deg] & [deg]  &[mag]& [mag] & [mag]    & [d]     &     & \\
  \hline
V01&251.84545&-1.92666&18.93&0.72&0.25&0.243185&EW&Y\\
V02&251.88581&-2.05289&18.31&1.18&0.12&0.252125&EW&N\\
V03&251.80287&-1.95726&17.86&0.72&0.08&0.210636&EW&Y\\
V04&251.90221&-1.91481&19.45&1.18&0.46&0.233990&EW&N\\
V05&251.79329&-1.93622&18.71&0.67&0.13&0.225502&EW&U\\
V06&251.83856&-1.91462&18.80&0.74&0.34&0.256196&EW&Y\\
V07&251.72871&-1.99055&16.65&1.06&0.07&0.257078&EW&N\\
V08&251.83855&-1.95674&16.48&0.35&0.07&0.435135&EW/Ell,BS&Y\\
V09&251.82010&-1.94197&17.12&0.77&0.03&0.444540&EW/Ell&U\\
V10&251.77274&-1.99331&19.06&0.71&0.34&4.595098&EA&Y\\
V11&251.81257&-1.95160&18.26&0.72&0.36&5.218589&EA&Y\\
V12&251.82194&-1.92852&16.09&0.37&0.10&1.025114&EA,BS&Y\\
V13&251.81159&-1.94571&17.07&0.47&0.26&0.734141&EA,BS&Y\\
V14&251.80630&-1.95057&17.58&0.54&0.62&0.463846&EA/EB,BS&Y\\
V15&251.80844&-1.94113&17.13&0.38&0.16&0.541709&EA,BS&U\\
V16&251.86936&-1.92680&17.30&0.83&0.45&2.500325&EA&N\\
V17&251.83190&-1.86607&20.00&1.30&0.82&2.279741&EA&U\\
V18&251.71477&-1.95946&16.46&0.90&0.65&17.47090&EA&N\\
V19&251.73301&-2.00912&19.64&0.83&0.97&-&EA&Y\\
V20&251.83532&-1.95287&19.72&0.75&0.62&-&EA&U\\
V21&251.80641&-1.94114&17.11&0.41&0.06&0.019636&SX,BS&Y\\
V22&251.84708&-1.92796&17.25&0.46&0.05&0.034364&SX,BS&Y\\
V23&251.82202&-1.96167&16.70&0.41&0.08&0.044294&SX,BS&Y\\
V24&251.80911&-1.89787&17.20&0.48&0.08&0.045042&SX,BS&Y\\
V25&251.76203&-1.96051&17.02&0.55&0.17&0.049034&SX,BS&Y\\
V26&251.80863&-1.99363&16.57&0.83&0.12&1.843644$^c$&Sp?&Y\\
V27&251.84428&-2.01776&17.89&1.12&0.44&11.36074$^c$&Sp?,RG?&N\\
V28&251.81727&-1.93884&13.85&1.00&0.04&0.949850$^c$&Sp?&Y\\
V29&251.80863&-1.99363&16.58&0.75&0.10&1.842144$^c$&Sp?&Y\\
V30&251.81196&-1.93522&16.39&0.50&0.03&0.400692&Ell?,BS&Y\\
V31&251.81356&-1.95517&17.61&0.44&0.08&0.405218&Ell?,BS &U\\
V32&251.80431&-1.95410&17.23&0.55&0.07&0.499084&Ell?&U\\
V33&251.81356&-1.95518&17.53&0.47&0.09&0.682552&Ell?,BS&Y\\
V34&251.80367&-1.96689&19.55&0.81&0.41&0.352252&Sp?&U\\
V35&251.80112&-1.95547&17.95&0.63&0.07&0.962294&Ell?&U\\
V36&251.77580&-1.95956&16.81&0.82&0.02&0.774844&Ell?&Y\\
  \hline
 \end{tabular}
\end{center}
{\footnotesize 
$^*$ We follow the naming convention of von Braun et al. (2002), who are the discoverers 
of the first two variables listed in the table. Their variable V1 is not the W Vir pulsator 
referred to as V1 by Clement et al. (1988).\\
$^a$EA - detached eclipsing binary, EB - close eclipsing binary, EW - contact eclipsing binary, 
SX~-~SX~Phe type pulsator, Sp - spotted variable, BS- blue straggler, 
Ell - ellipsoidal variable, RG - red giant.\\ 
$^b$Y - member, N - nonmember, U - no data or data ambiguous\\
$^c$For the 2001 season.}
\end{table}

\clearpage

\begin{subfigures}
\begin{figure}
   \centerline{\includegraphics[width=0.95\textwidth,
               bb = 42 26 528 767, clip]{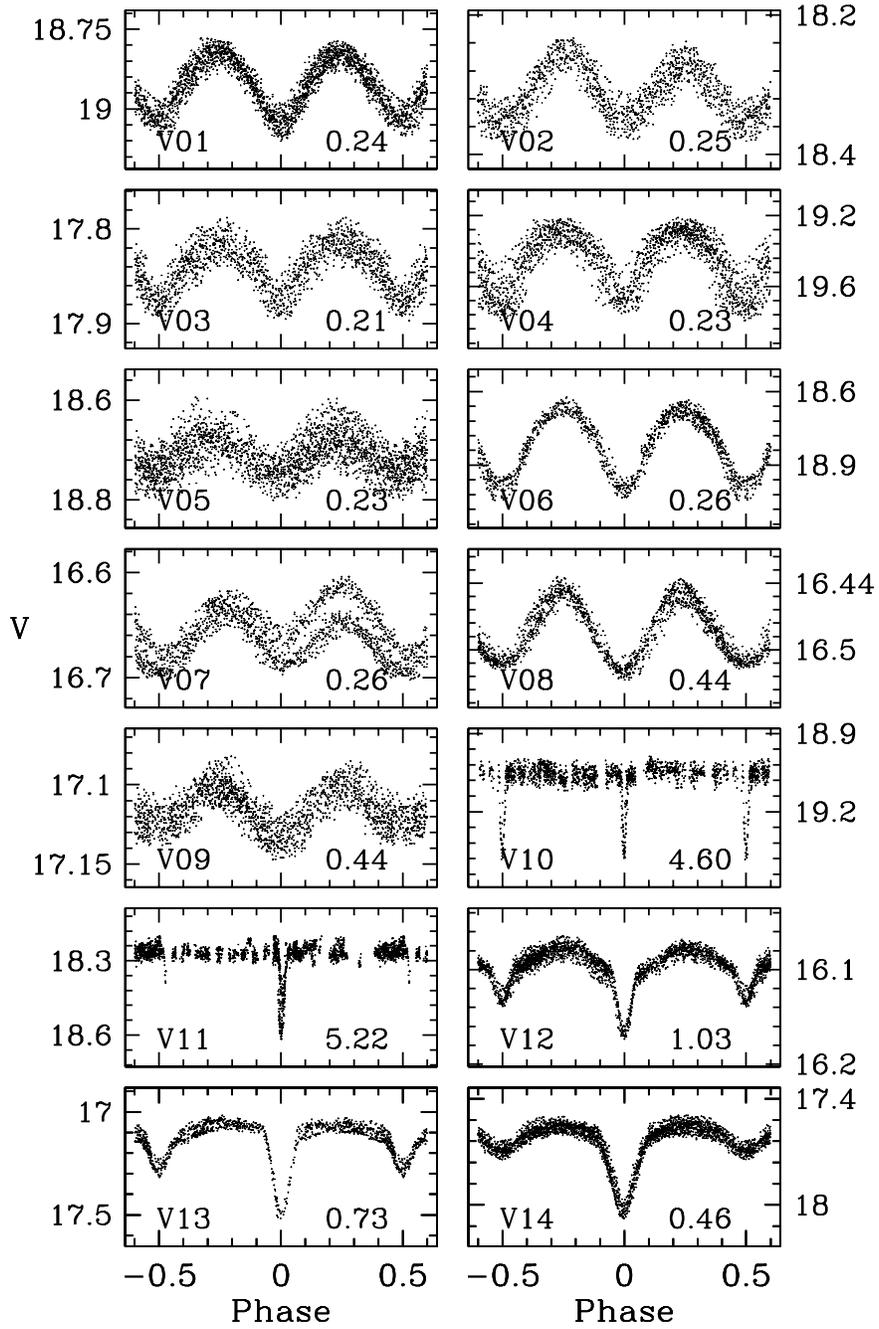}}
   \caption{Phased $V$ curves of variables detected in the field of
    M12. Inserted labels give star ID and period in days.
    \label{fig:curves1}}
\end{figure}

\begin{figure}
   \centerline{\includegraphics[width=0.95\textwidth,
       bb = 42 26 528 765, clip]{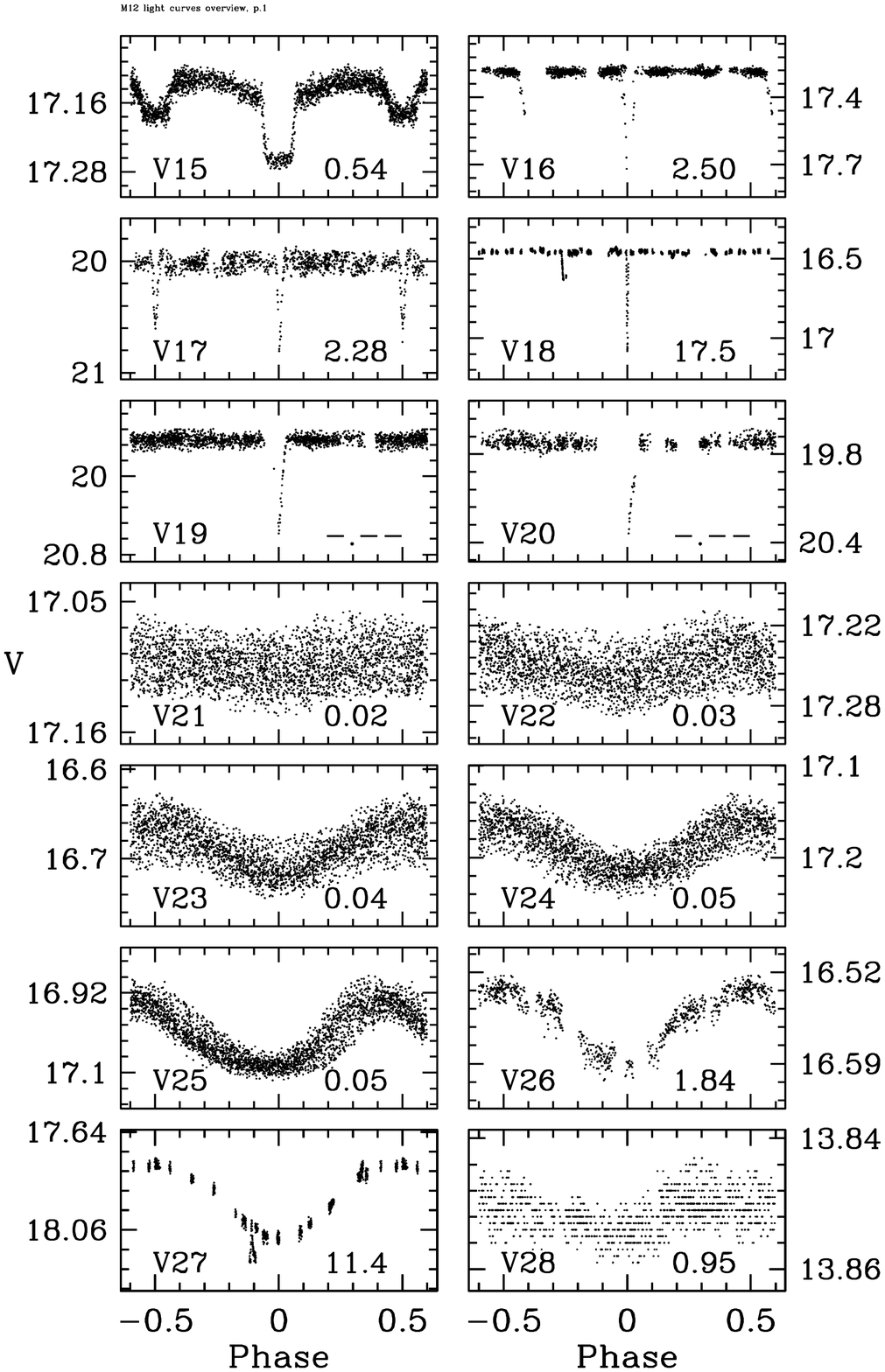}}
   \caption{Continuation of Fig. \ref{fig:curves1}.  
    \label{fig:curves2}}
\end{figure}

\begin{figure}
   \centerline{\includegraphics[width=0.95\textwidth,
       bb = 42 328 528 765, clip]{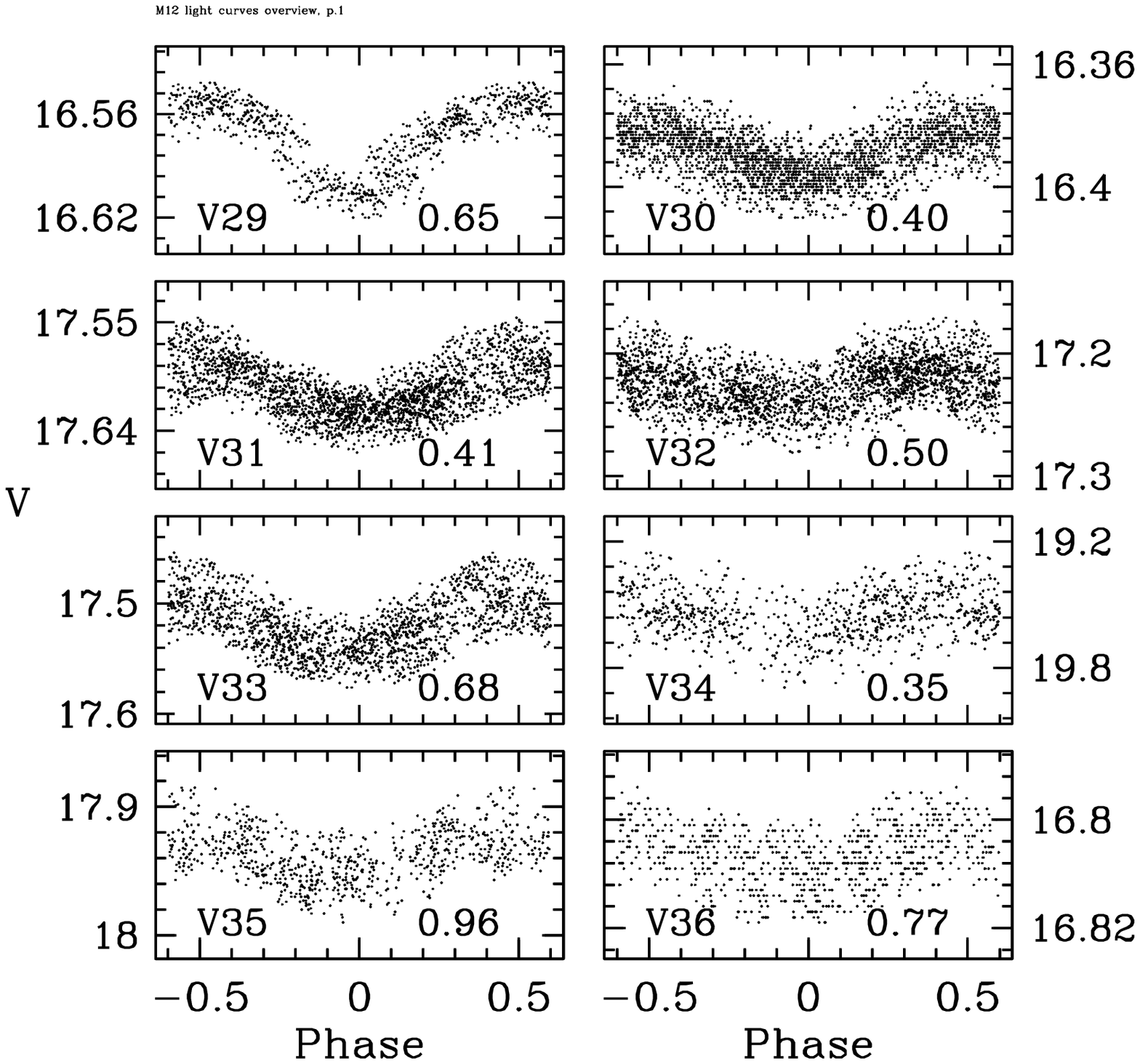}}
   \caption{Continuation of Fig. \ref{fig:curves1}.  
    \label{fig:curves3}}
\end{figure}
\end{subfigures}

\clearpage

\end{document}